\documentclass[12pt,notitlepage]{article}

\usepackage{amsmath,amssymb}
\usepackage{graphicx}

\usepackage[normalem]{ulem}

\usepackage[square,numbers]{natbib}
\newcommand\Mycite[1]{%
	\citeauthor{#1}~(\citeyear{#1})}

\newcommand\Mycitet[1]{%
	\citeauthor{#1}~\citeyear{#1}}

\usepackage{color}

\def\ignore#1{{}}

\newcounter{sxn}

\newcounter{axn}

\newdimen\mybaselineskip
\newcommand{\beeq}{\begin{equation}}
\newcommand{\eneq}{\end{equation}}
\newcommand{\beqn}{\begin{eqnarray}}
\newcommand{\eeqn}{\end{eqnarray}}

\newcommand{\ba}{\begin{array}}
\newcommand{\ea}{\end{array}}

\newcommand{\be}{\begin{equation}}
\newcommand{\ee}{\end{equation}}
\newcommand{\bea}{\begin{eqnarray}}
\newcommand{\eea}{\end{eqnarray}}
\newcommand{\beal}{\setcounter{letter}{1} \begin{eqnarray}}
\newcommand{\eeal}{\addtocounter{equation}{1} \end{eqnarray}}

\newcommand{\larrow}{\,\,\,\,\hbox to 30pt{\rightarrowfill}
\,\,\,\,}
\newcommand{\slarrow}{\,\,\,\hbox to 20pt{\rightarrowfill}
\,\,\,}

\def\la{\raise.16ex\hbox{$\langle$}\lower.16ex\hbox{}  }
\def\ra{\, \raise.16ex\hbox{$\rangle$}\lower.16ex\hbox{} }

\def\psibar{ \psi \kern-.65em\raise.6em\hbox{$-$} \lower.6em\hbox{} }
\def\psibarb{ \psi \kern-.65em\raise.6em\hbox{$-$}  }

\begin{document}

\thispagestyle{empty}

\begin{center}  
	{\LARGE \bf  Erratum: ``Spacetime Metrics and Ringdown Waveforms for Galactic Black Holes Surrounded by a Dark Matter Spike''\\ \textcolor{blue}{(The Astrophysical Journal, 940:33 (10pp), 2022 November 20)}}
	\vspace{1cm}
	
	{\bf  Ramin G.~Daghigh$^{1}$ and Gabor Kunstatter$^2$}
\end{center}

\centerline{\small \it $^1$ Natural Sciences Department, Metropolitan State University, Saint Paul, Minnesota, USA 55106}
\vskip 0 cm
\centerline{}

\centerline{\small \it $^2$ Physics Department, University of Winnipeg, Winnipeg, MB Canada R3B 2E9}
\vskip 0 cm
\centerline{} 

\vspace{0cm}

In \cite{DK_AppJ}, we used the Tolman-Oppenheimer-Volkoff (TOV) equations to calculate the effects of an isotropic dark matter spike on the ringdown waveform and shadow of the supermassive black hole at the core of M87. 

The assumption of isotropy in this context, also used in earlier work\cite{Nampalliwar}, is  suspect in the case of non-interacting dust since near the photon sphere, where the motion is highly relativistic, non-zero radial pressure necessarily implies a flow of matter into the black hole and renders the solution non-static\footnote{The authors are grateful to Don Page for making us aware of this argument.}. While the isotropic TOV equations  used in \cite{DK_AppJ} imply that the radial pressure has negligible impact on the spacetime geometry for physical parameters relevant to galactic black holes, this result is not strictly justified for the non-interacting dust. However, the validity of neglecting the pressure has been confirmed in a separate calculation that starts from non-isotropic pressure\cite{DK2024}. The assumption of isotropy can be viable in certain regions of the dark matter halo\cite{Navarro, Ludlow} and in certain scenarios such as self-interacting dark matter spikes\cite{Saxton}.

While the numerical results and overall conclusions about observability of the dark matter spikes were qualitatively correct,  the calculation was done in a frame in which the metric of the vacuum inside the spike was Schwarzschild. In reality, the effect is measured in the frame of an asymptotic observer and was therefore underestimated. There exists an overall redshift for the asymptotic observer which can increase the effect by anywhere from 3\% to 40\%, depending on the magnitude of the density of the spike. Details of the redshift calculation can be found in \cite{DK2024}. 
We also  note  a typographical error in Eqs.\ (30) and (31) of \cite{DK_AppJ}. The term $\left( 1+r_0/r \right)$ should be $\left( 1+r/r_0 \right)$.  The numerical results were not affected.\\[10pt]

\leftline{\bf Acknowledgments}
We thank Andrei Frolov and Don Page for helpful discussions.  G.K.\ gratefully acknowledges that this research was supported in part by a Discovery Grant from the Natural Sciences and Engineering Research Council of
Canada.

\newpage

\begin{center}  
{\LARGE \bf   Spacetime metrics and ringdown waveforms for galactic black holes surrounded by a dark matter spike}

\vspace{1cm}

{\bf  Ramin G.~Daghigh$^{1}$ and Gabor Kunstatter$^2$}
\end{center}

\centerline{\small \it $^1$ Natural Sciences Department, Metropolitan State University, Saint Paul, Minnesota, USA 55106}
\vskip 0 cm
\centerline{}

\centerline{\small \it $^2$ Physics Department, University of Winnipeg, Winnipeg, MB Canada R3B 2E9}
\vskip 0 cm
\centerline{} 

\vspace{1cm}

\begin{abstract}
Theoretical models suggest the existence of a dark matter spike surrounding the supermassive black holes at the core of galaxies. The spike density is thought to obey a power law that starts at a few times the black hole horizon radius and extends to a distance, $R_\text{sp}$, of the order of a kiloparsec. We use the Tolman-Oppenheimer-Volkoff equations to construct the spacetime metric representing a black hole surrounded by such a dark matter spike.  We consider the dark matter to be a perfect fluid, but make no other assumption about its nature.  The assumed power law density provides in principle three parameters with which to work: the power law exponent $\gamma_\text{sp}$, the external radius $R_\text{sp}$, and the spike density $\rho_\text{DM}^\text{sp}$ at $R_\text{sp}$.  These in turn determine the total mass of the spike. We focus on Sagittarius A* and M87 for which some theoretical and observational bounds exist on the  spike parameters. Using these bounds in conjunction with the metric obtained from the Tolman-Oppenheimer-Volkoff equations,  we investigate the possibility of detecting the dark matter spikes surrounding these black holes via the gravitational waves emitted at the ringdown phase of  black hole perturbations.  Our results suggest that if the spike to black hole mass ratio is roughly constant, greater mass black holes require relatively smaller spike densities to yield potentially observable signals. 
We find that is unlikely for the spike in M87 to be detected via the ringdown waveform with currently available techniques unless its mass is roughly an order of magnitude larger than existing observational estimates.
However, given that the signal increases with black hole mass, spikes might be observable for more massive galactic black holes in the not too distant future.
\end{abstract}

\newpage

\section{Introduction}
There is solid observational evidence from the spiral galaxy rotation curves and mass-luminosity ratios of elliptical galaxies that a dark matter (DM) halo encompasses every galaxy and fills the intergalactic medium.   The shape of the DM density profile of this halo is less known, but could play a vital role in determining the geometry of spacetime near the galactic center.
Multiple models exist for the spacetime metric around a static and spherically symmetric black hole with a DM halo based on the Newtonian approximation, including  the Navarro-Frenk-White (NFW) profile (\Mycitet{NFW1}; \Mycitet{NFW2}) and Burkert–Salucci profile (\Mycitet{Burkart1}; \Mycitet{Burkart2}).  See also  \Mycite{BH-DMhalo1}, \Mycite{BH-DMhalo2}, \Mycite{BH-DMhalo3}, \Mycite{BH-DMhalo4}, and \Mycite{BH-DMhalo5}.    

It has  been argued that the adiabatic growth of a black hole immersed in cold DM can lead to the formation of high density regions of DM known as ``spikes" around supermassive (\Mycitet{DMspikeformation1}; \Mycitet{GondoloSilk}; \Mycitet{DMspikeformation3}) and intermediate
mass (\Mycitet{DMspikeformation4}; \Mycitet{DMspikeformation5}; \Mycitet{DMspikeformation6}) black holes.   The first DM spike model was described by \Mycite{GondoloSilk}, who proposed a power law density distribution for the DM.   \Mycite{Sadeghian} included general relativistic corrections to the model of Gondolo and Silk (G-S) and found that the  density distribution of the DM around  Schwarzschild black holes would begin at around twice the horizon radius instead of four times as proposed initially by G-S. \Mycite{Sadeghian} also found that the peak density of the DM spike was $15$ percent  higher as compared to the Newtonian approximation used by G-S. This suggests that the DM spike may have
important implications for observations.  

The relativistic corrections to the spacetime metric for a black hole surrounded by a DM spike was constructed in \Mycite{BH-DMspike1} and \Mycite{BH-DMspike2} starting from the power law density profile proposed by G-S.  \Mycite{BH-DMspike1} assume $g_{tt}=-g_{rr}^{-1}$ and use perturbative approximations while \Mycite{BH-DMspike2} calculate the metric components to leading order of spike density at the outside edge.


The prospects of detecting the DM spike using gravitational waves have been investigated in \Mycite{DMspikeGW1}, \Mycite{DMspikeGW2}, \Mycite{DMspikeGW3}, \Mycite{DMspikeGW4}, \Mycite{DMspikeGW5},  and \Mycite{DMspikeGW6}, which focus mainly on the waveform of extreme and/or intermediate mass ratio inspirals. The ringdown waveform and quasinormal modes (QNMs) of a black hole in a cold DM halo were studied in \Mycite{QNMhalo1}, \Mycite{QNMhalo2}, \Mycite{QNMhalo3}, and \Mycite{QNMhalo4}.  In the present paper, we focus on the impact of DM spikes on the metric, ringdown waveforms, and QNMs of supermassive black holes, specifically those at the center of Sagittarius A* (Sgr A*) and M87.  The hope is to  discover signals detectable at least in principle.  There may not exist a mechanism, such as  extreme mass ratio inspiral or  galaxy collision, to emit detectable gravitational waves from these two galaxies.  However, this issue is not a  deterrent since it is inevitable that some of the many galaxies in the universe have the right conditions to produce waves that can be detected by the current or next generation of gravitational wave experiments.

Following previous work, we assume a power law density for the DM spike.  The assumed power law density provides three parameters with which to work, namely, the power law exponent $\gamma_\text{sp}$, as well as the radius $R_\text{sp}$ and spike density $\rho_\text{DM}^\text{sp}$ at one of the spike boundaries normally taken to be  the outside edge. These in turn determine the total mass of the spike. If one assumes a power law density profile,  the pressure and metric components must be derived from the full Tolman-Oppenheimer-Volkoff (TOV) equations.
The main features that emerge from our analysis are:
\begin{itemize}
	\item It turns out that the pressure is small enough to be neglected in the TOV equations.
	This allows us to obtain a self-consistent\footnote{By self-consistent we mean that we use the resulting metric in the corresponding TOV equation to verify the smallness of the pressure.} analytic expression for the metric components.
	\footnote{We find that our analytic expression for the $tt$ metric component differs substantially from those of   Nampilliwar et al. in our region of interest.}
	\item For Sgr A*,  the parameters have to be pushed well beyond the accepted ranges in order to produce significant differences from the Schwarzschild ringdown waveform.
	\item For M87, the parameters are less known, but there is an observational  bound on the total mass within $50$ kpc of the center, which in turn provides an upper bound on the spike mass.  We show that there exist values for the spike parameters, consistent qualitatively with those of Sgr A* and producing a total spike mass within the bound for M87, that 
	significantly enhances the differences from the Schwarzschild ringdown waveform in comparison to  Sgr A*.
	\item Assuming that the ratio of the DM spike mass grows roughly linearly with the black hole mass, the relative effect on the ringdown waveforms increases with total mass.
	\item { One might wonder about the impact of the regular mass, in the region near the black hole, on the ringdown waveform.  The lowest estimate for the radius of the galactic bulge, surrounding Sgr A*, is approximately  $2$ kpc and the highest estimate for the bulge mass is $2\times 10^{10}$ solar masses (see \Mycitet{bulge-SgrA}).  Using these values, we can find an upper bound for the average density of the bulge, which is approximately $4.0\times 10^{-23}$ g/cm$^3$.  This is an order of magnitude less than the average density of the spike, surrounding Sgr A*, in the region $r<R_\text{sp}$.  In addition, the dark matter density at the inner edge of the spike, i.e. near the black hole horizon, is approximately $10^{19}$ times higher than the average spike density.    As can be seen in Figs.\ \ref{fig-g1MW} and \ref{fig-g0MW} of this paper, the effective potential that determines the ringdown waveform drops rapidly to zero for large $r$. Therefore, in the region that produces the dominant effect on the ringdown waveform, one can safely ignore the bulge.  We assume this is also true for M87, for which less is known about the mass distribution.}	
\end{itemize}

We structure the paper as follows. In Sec.\ \ref{Sec:TOV}, we set up the problem by reviewing the relevant TOV equations and associated boundary conditions. We then solve for the pressure and metric assuming a power law density for the DM spike.   In Sec.\ \ref{Sec:WE}, we briefly review the wave equation for scalar field perturbations in the black hole background.  Sec. \ref{Sec:SgrBH} calculates the ringdown waveform and the lowest QNM for the multipole number $l=2$ for the SgrA* DM spike, while Sec. \ref{Sec:M87} does the same for M87. We conclude in Sec. \ref{Sec:conclusions} with a summary of the results.

\section{Solving the Tolman-Oppenheimer-Volkoff Equations}
\label{Sec:TOV}

We start with the most general $4$-D spherically symmetric static metric (up to coordinate transformations)
\beeq
ds^2 = - e^{\mu(r)}dt^2 + \left(1-\frac{2M(r)}{r}\right)^{-1}dr^2 + r^2 d\Omega^2,
\label{eq:GeneralMetric}
\eneq
and assume a perfect fluid stress tensor for the DM spike
\bea
T^\mu_\nu = \hbox{diag}(\rho(r), p(r), p(r), p(r)).
\eea
This yields the TOV equations (\Mycitet{Carroll}) in the spike region:
\bea
G_{tt}&=0& \implies \frac{dM(r)}{dr} = 4\pi r^2 \rho(r)
\label{eq:Gtt}\\
G_{rr}&=0& \implies \frac{d\mu(r)}{dr} = 2\frac{M(r)+4\pi r^3 p(r)}{r\left[r-2M(r)\right]}
\label{eq:Grr}\\
\partial_\nu T^{r\nu} &=& 0 \implies \frac{dp(r)}{dr}=-[\rho(r)+p(r)]\frac{M(r)+4\pi r^3p(r)}{r\left[r-2M(r)\right]}
\label{eq:MomentumConservation}
\eea
We have three equations in four unknowns $[\mu(r), M(r), \rho(r), p(r)]$ so they need to be supplemented by a fourth equation. Normally this is taken to be the equation of state relating $\rho$ to $p$. In the present case, we wish to assume a particular density profile for the DM spike, which provides the extra equation. There is no freedom left to specify the equation of state. We show, however, that one can assume the pressure is negligible when solving for $\mu(r)$ in Eq.\ (\ref{eq:Grr}).

We now introduce the density profile for the DM spike.  Given a black hole with a mass $M_{\text{BH}}$ at a galactic center surrounded by a DM halo with an initial power law density profile 
\begin{equation}
	\rho_{\text{DM}}(r)\simeq \rho_0 \left( \frac{r_0}{r} \right)^\gamma,
	\label{}
\end{equation}
where $\gamma$ is the power law index and $\rho_0$ and $r_0$ are the halo parameters,   it has been shown  (\Mycitet{GondoloSilk}) that a DM spike will form adiabatically with a density profile
\begin{equation}
	\rho_{\text{DM}}^\text{sp}(r)\simeq \rho_{\text{sp}} \left( \frac{R_\text{sp}}{r} \right)^{\gamma_\text{sp}}=\rho_{b} \left( \frac{r_\text{b}}{r}\right)^{\gamma_{\text{sp}}},
	\label{eq-SpikeDensity}
\end{equation}
where 
\begin{equation}
	\rho_{\text{sp}}= \rho_0 \left( \frac{R_\text{sp}}{r_0} \right)^{-\gamma},~R_{\text{sp}}=\alpha_\gamma r_0 \left( \frac{M_{\text{BH}}}{\rho_0 r_0^3}\right)^{\frac{1}{3-\gamma}}, ~ \text{and}~\gamma_{\text{sp}}=\frac{9-2\gamma}{4-\gamma}.
	\label{eq-spike-parameters}
\end{equation}
Here, $\rho_\text{sp}$ and $R_{\text{sp}}$ are the density and radius of the spike, respectively, at the outer edge.  Instead of  $\rho_\text{sp}$ and  $R_{\text{sp}}$, one can  use $\rho_{\text{b}}$ and  $r_{\text{b}}$, which are the density and radius of the spike at its inner edge.  In this paper, we will use the former.  
Using Eq.\ (\ref{eq-spike-parameters}), one can show that $\alpha_\gamma$ is related to the spike parameters according to
\begin{equation}
	\alpha_\gamma= \left(\frac{\rho_{\text{sp}} R_{\text{sp}}^3}{M_{\text{BH}}}\right)^{\frac{1}{3-\gamma}}.
	\label{eq-alpha-gamma}
\end{equation}


We  can substitute the DM spike density profile (\ref{eq-SpikeDensity}) into Eq.\ (\ref{eq:Gtt}) to solve for the metric or mass function $M(r)$ in the spike region ($ r_\text{b}\le r \le R_{\text{sp}}$). The overall mass function at different regions can be summarized as (\Mycitet{BH-DMspike2})
\begin{equation}
	M(r)=
	\left\{ {\begin{array}{cc}
			M_{\text{BH}} & r \le r_\text{b} \\
			M_{\text{BH}}+M_{\text{DM}}^{\text{sp}}(r) & r_\text{b}\le r \le R_{\text{sp}} \\
			M_{\text{BH}}+M_{\text{DM}}  & r > R_{\text{sp}} \\
	\end{array} } \right. 
	\label{}
\end{equation}
where
\bea
M_{\text{DM}}^{\text{sp}}(r)=\frac{4 \pi \rho_{\text{sp}}}{3-\gamma_{\text{sp}}}\left[r^3 \left(\frac{R_{\text{sp}}}{r}\right)^{\gamma_{\text{sp}}}-  r_\text{b}^3 \left(\frac{R_{\text{sp}}}{r_\text{b}}\right)^{\gamma_{\text{sp}}} \right]
\label{eq-SpikeMass}
\eea
is the mass function of the DM spike. $M_\text{DM}$ is the combined mass of the spike and the DM halo surrounding the spike within a radius $r > R_{\text{sp}}$.  The impact of this region on ringdown waveforms is negligible.  See Sections \ref{Sec:SgrBH} and \ref{Sec:M87} for more details.
Here, we use geometrized unit system where $c=G=1$.


{ Note that the total mass of the spike, $M_\text{total}^\text{sp}=M_{\text{DM}}^{\text{sp}}(R_\text{sp})$, can be increased by increasing $R_\text{sp}$, $\rho_\text{sp}$, or both. The total mass is  proportional to $\rho_\text{sp}$ and  $R_\text{sp}^3$ so that, according to Eq.\ (\ref{eq-alpha-gamma}), increasing the mass of the spike requires $\alpha_\gamma$ to increase. In this paper, we increase $M_\text{total}^\text{sp}$ by increasing $\rho_{sp}$ and keeping $R_{sp}$ fixed.

Given the large variety of different parameters used to describe the spike and halo in the literature, we summarize our general framework as follows: In addition to the mass $M_\text{BH}$ of the black hole, four parameters are required. We take these to be the exponent $\gamma_\text{sp}$, the location $r_\text{b}$ of the inner edge of the spike, the density $\rho_\text{sp}$ at the outer edge, and $\alpha_\gamma$. These are sufficient to determine all other spike parameters, including the location $R_\text{sp}$ of the outer edge via Eq.\ (\ref{eq-alpha-gamma}) and the total mass of the spike via Eq.\ (\ref{eq-SpikeMass}). Experiment provides an upper bound on the total mass of the spike plus halo, but not on the other parameters.\footnote{Note that  $\rho_\text{sp}$ can be matched to the halo density at $R_\text{sp}$.}}

Next, we need to solve Eq.\ (\ref{eq:MomentumConservation}) for $p(r)$. This is not possible analytically, but we have solved it numerically using the built-in {\em Mathematica} commands for solving differential equations. It turns out that the term $4\pi r^3 p(r)$ can be neglected compared to $M(r)$ in Eq.\ (\ref{eq:Grr}). Using this approximation, Eq.\ (\ref{eq:Grr}) can be written as
\bea
\frac{d\mu(r)}{dr} = -\frac{1}{r} +\frac{1}{\left[r-2(M_{\text{BH}}+a r^{3-\gamma_{\text{sp}}}-b)\right]},
\label{eq:alpha}
\eea
where $a=\frac{4 \pi \rho_{\text{sp}}}{3-\gamma_{\text{sp}}}R_{\text{sp}}^{\gamma_{\text{sp}}}$ and $b=\frac{4 \pi \rho_{\text{sp}} r_\text{b}^3}{3-\gamma_{\text{sp}}} \left(\frac{R_{\text{sp}}}{r_\text{b}}\right)^{\gamma_{\text{sp}}}$.  

We first take the case where $\gamma_{\text{sp}}=7/3$ ($\gamma=1$).\footnote{For more discussion on the values of spike parameters, see \Mycite{Feng}.}  We can now integrate Eq.\ (\ref{eq:alpha}) to get
\begin{eqnarray}
	\mu(r) &=& -\int  \frac{3dy}{y} +\int \frac{3y^2 dy}{\left[y^3-2(M_{\text{BH}}+a y^{2}-b)\right]}+C \nonumber \\
	&=& -\int  \frac{3dy}{y} +\frac{3 y_0^2}{y_1 y_2+y_0(y_0-y_1-y_2)}\int \frac{ dy}{y-y_0}\nonumber \\
	&& +\frac{3 }{y_1 y_2+y_0(y_0-y_1-y_2)}\int \frac{(y_0 y_1 y_2-y_0 y_1 y-y_0 y_2 y+y_1 y_2 y) dy}{y^2-(y_1+y_2)y+y_1 y_2},
	\label{eq: alpha7:3}
\end{eqnarray}
where we have used the change of variable $y=r^{1/3}$.  Here, $y_0$ is the real root of the equation $y^3-2(M_{\text{BH}}+a y^{2}-b)$ and $y_1$ and $y_2$ are the two complex conjugate roots. After integration, the final result for the metric function, $f(r)=e^{\mu(r)}$, is
\begin{eqnarray}
	f(r) &=&\left(1-\frac{2M_{\text{BH}}}{r_\text{b}}\right)\frac{r_\text{b}}{r} \left(\frac{r^{1/3}-y_0}{r_\text{b}^{1/3}-y_0}\right)^{\frac{3y_0^2}{y_1 y_2+y_0(y_0-y_1-y_2)}} \nonumber \\
	&&\left(\frac{r^{2/3}-(y_1+y_2)r^{1/3}+y_1y_2}{r_\text{b}^{2/3}-(y_1+y_2)r_\text{b}^{1/3}+y_1y_2}\right)^{\frac{3[y_1 y_2-y_0(y_1+y_2)]}{2[y_1 y_2+y_0(y_0-y_1-y_2)]}} \nonumber \\
	&&e^{   \frac{3[2 y_0 y_1 y_2-y_0(y_1+y_2)^2+y_1 y_2 (y_1+y_2)]}{\sqrt{4y_1y_2-(y_1+y_2)^2}[y_1 y_2+y_0(y_0-y_1-y_2)]} \left(\arctan{\frac{2r^{1/3}-y_1-y_2}{\sqrt{4y_1y_2-(y_1+y_2)^2}}}-\arctan{\frac{2r_\text{b}^{1/3}-y_1-y_2}{\sqrt{4y_1y_2-(y_1+y_2)^2}}}\right)}.
	\label{eq:f-7over3}
\end{eqnarray}
We have chosen the constant of integration, $C$, so that $f(r_\text{b}) =1-2M_{\text{BH}}/r_\text{b}$.

\begin{figure}[th!]
	\begin{center}
		\includegraphics[height=4.8cm]{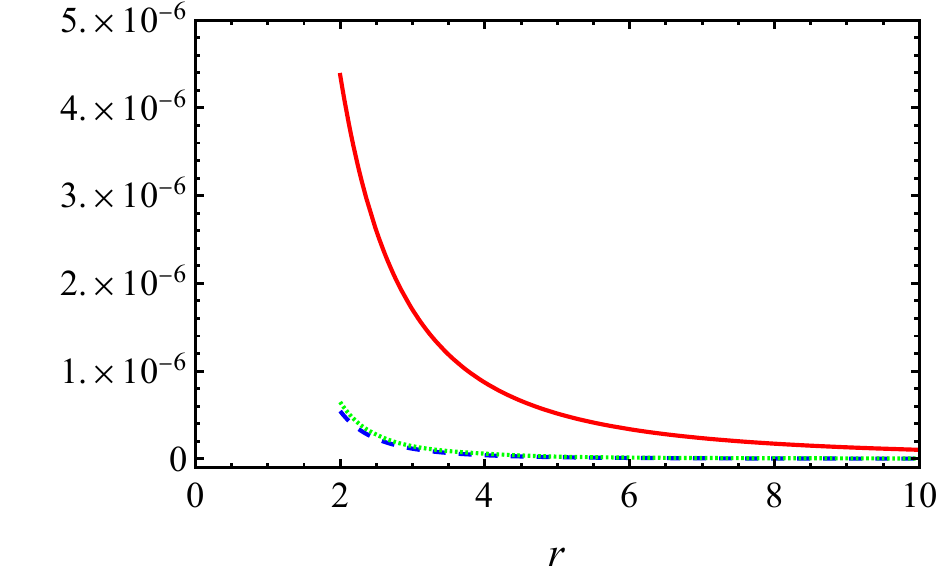}
		\includegraphics[height=4.8cm]{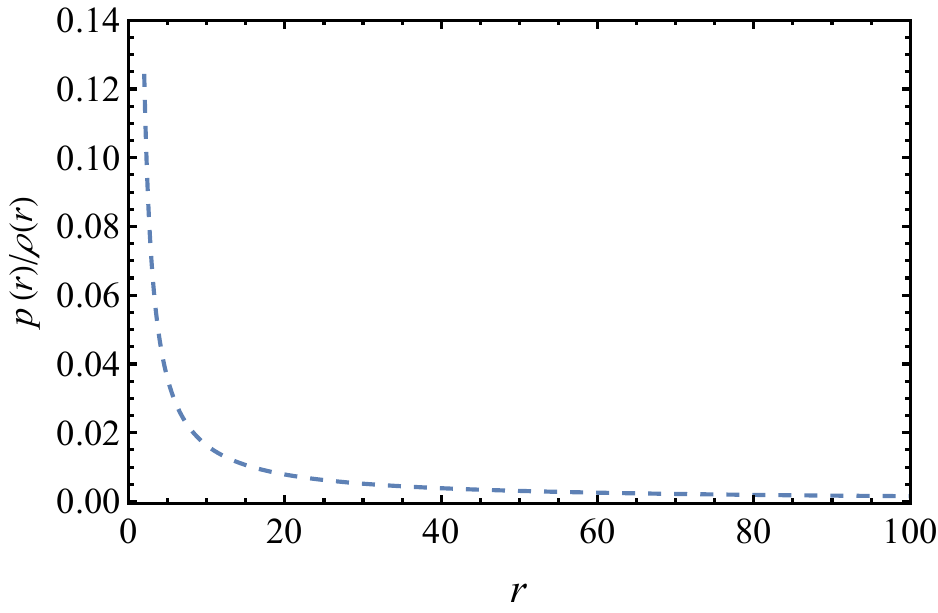}
	\end{center}
	\vspace{-0.7cm}
	\caption{\footnotesize On the left, we plot the DM spike pressure $p(r)$ obtained analytically using Eq.\ (\ref{eq: p-analytic}) in dashed blue and  numerically using Eq.\ (\ref{eq:MomentumConservation}) in dotted green. For comparison, we include the DM spike density $\rho_{\text{DM}}^{\text{sp}}(r)$ in solid red.   We take $\gamma_{\text{sp}}=7/3$, $r_\text{b}=2 r_{\text{BH}}$, and use the Sgr A* data where $R_{sp}=0.235$ kpc and $\rho_{sp}=6.7 \times 10^{-22}$ g cm$^{-3}$ ($\approx 8$ times the expected value).  On the right, for the same spike parameters, we plot pressure [from Eq.\ (\ref{eq: p-analytic})] divided by the density of the DM spike to show that pressure stays negligible everywhere.  All our variables are expressed in terms of black hole parameters ($r_\text{BH}$ and $\rho_\text{BH}$ defined in Sec.\ \ref{Sec:SgrBH}.) }
	\label{fig: p-rho}
\end{figure}

\begin{figure}[th!]
	\begin{center}
		\includegraphics[height=4.8cm]{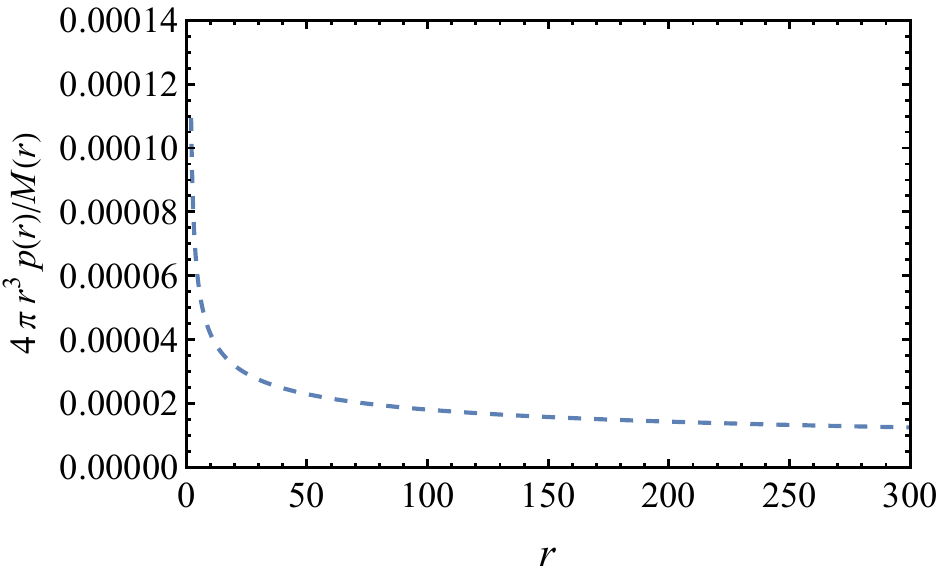}
	\end{center}
	\vspace{-0.7cm}
	\caption{\footnotesize We plot  $4 \pi r^3 p(r)/M(r)$ in the DM spike region to show that neglecting pressure in Eq.\ (\ref{eq:Grr}) is valid.  We take $\gamma_{\text{sp}}=7/3$, $r_\text{b}=2 r_{\text{BH}}$, and use the Sgr A* data where $R_{sp}=0.235$ kpc and $\rho_{sp}=6.7 \times 10^{-22}$ g cm$^{-3}$ ($\approx 8$ times the expected value). The radius $r$ is in units of the black hole horizon radius.  }
	\label{fig: p-M}
\end{figure}

We also want to show that pressure is negligible in the spike region.  Assuming $p(r)/\rho(r) \ll 1$ and  $4 \pi r^3 p(r)/M(r) \ll 1$, one can rewrite Eq.\ (\ref{eq:MomentumConservation}) as
\bea
 \frac{dp(r)}{dr}=-\rho_\text{DM}^\text{sp}(r)\frac{M_\text{DM}^\text{sp}(r)}{r\left[r-2M_\text{DM}^\text{sp}(r)\right]},
\label{eq: p'}
\eea
where we have replaced $\rho$ and $M$ with the spike parameters.
We then find the approximate pressure by integrating Eq.\ (\ref{eq: p'}),
\begin{equation}
	p(r)= -\frac{1}{2} \rho_{\text{sp}}R_{\text{sp}}^{7/3}\left\{-\int_{\infty}^{y^3}  \frac{3dy}{y^8} +\int_{\infty}^{y^3} \frac{3 dy}{y^5\left[y^3-2(M_{\text{BH}}+a y^{2}-b)\right]}\right\}. 
	\label{eq: p-analytic}
\end{equation}
For this approximation to be valid/consistent, the pressure from the above equation should satisfy the same conditions, i.e.\ $p(r)/\rho(r) \ll 1$ and  $4 \pi r^3 p(r)/M(r) \ll 1$.
The integration in Eq.\ (\ref{eq: p-analytic}) can be handled analytically by writing the integrand as a sum of terms with minimal denominators similar to what we do in Eq.\ (\ref{eq: alpha7:3}).  We plot the  pressure, density, and the ratio of the two as a function of the radial coordinate, in Figure \ref{fig: p-rho}, to show $p(r) \ll \rho_{\text{DM}}^{\text{sp}}(r)$.  In Figure \ref{fig: p-rho}, we also plot the DM spike pressure obtained numerically, using built-in {\em Mathematica} commands for differential equations, by solving Eq.\ (\ref{eq:MomentumConservation}) with no approximation.  Our numerical and analytical solutions are more or less the same.
In Figure \ref{fig: p-M}, we  plot  $4 \pi r^3 p(r)/M(r)$ as a function of the radial coordinate in the spike region to show that this term is also negligibly small.   Therefore, the spike pressure can be ignored in the TOV equations.  

We can also consider the case where $\gamma_{\text{sp}}=9/4$ ($\gamma=0$).  We  integrate Eq.\ (\ref{eq:alpha}) to get
\bea
\mu(r) = -\int  \frac{dr}{r} +4\int \frac{y^3 dy}{\left[y^4-2(M_{\text{BH}}+a y^{3}-b)\right]}+C,
\label{}
\eea
where we have used the change of variable $y=r^{1/4}$.  This integration can be handled analytically by writing the integrand as a sum of terms with minimal denominators similar to what we do in Eq.\ (\ref{eq: alpha7:3}). The final result for the metric function, $f(r)=e^{\mu(r)}$, is
\begin{eqnarray}
	f(r) &=&\left(1-\frac{2M_{\text{BH}}}{r_\text{b}}\right)\frac{r_\text{b}}{r} \left(\frac{r^{1/4}-y_0}{r_\text{b}^{1/4}-y_0}\right)^{\frac{4y_0^3}{(y_1 y_2+y_0(y_0-y_1-y_2))(y_0-y_3)}} \left(\frac{r^{1/4}-y_3}{r_\text{b}^{1/4}-y_3}\right)^{\frac{4y_3^3}{(y_1 y_2+y_3(y_3-y_1-y_2))(y_0-y_3)}} \nonumber \\
	&&\left(\frac{r^{1/2}-(y_1+y_2)r^{1/4}+y_1y_2}{r_\text{b}^{1/2}-(y_1+y_2)r_\text{b}^{1/4}+y_1y_2}\right)^{\frac{2[y_1^2 y_2^2+y_0y_3(y_1+y_2)^2-y_0y_1y_2y_3-y_1y_2(y_0+y_3)(y_1+y_2)]}{[y_1 y_2+y_0(y_0-y_1-y_2)][y_1 y_2+y_3(y_3-y_1-y_2)]}} \nonumber \\
	&&\exp\left\{ 4  {{\frac{y_0y_3(y_1+y_2)^3-y_1y_2(y_0+y_3)(y_1+y_2)^2+2y_1^2 y_2^2 (y_0+y_3)-y_1y_2(3y_0y_3-y_1y_2)(y_1+y_2) }{\sqrt{4y_1y_2-(y_1+y_2)^2}[y_1 y_2+y_0(y_0-y_1-y_2)][y_1 y_2+y_3(y_3-y_1-y_2)]}}} \right. \nonumber \\ 
	&&\left. \left(\arctan{\frac{2r^{1/4}-y_1-y_2}{\sqrt{4y_1y_2-(y_1+y_2)^2}}}-\arctan{\frac{2r_\text{b}^{1/4}-y_1-y_2}{\sqrt{4y_1y_2-(y_1+y_2)^2}}}\right)\right\},
	\label{}
\end{eqnarray}
where $y_0$ and $y_3$ are the real roots of the equation $y^4-2(M_{\text{BH}}+a y^{3}-b)$ and $y_1$ and $y_2$ are the two complex conjugate roots.  We have chosen the constant of integration, $C$, so that $f(r_\text{b}) =1-2M_{\text{BH}}/r_\text{b}$.

\section{Wave Equation}
\label{Sec:WE}

We wish to investigate the ringdown waveform emitted from a black hole surrounded by a DM spike. For simplicity, we look at scalar perturbations with the assumption that the graviton modes will have similar behavior.  {This assumption is based on the similarity of the Regge-Wheeler potential for scalar and gravitational perturbations.  See the explicit form of the potentials provided in, for example, \Mycite{Leaver}.}
A massless scalar field in the background of a black hole spacetime obeys the Klein-Gordon equation
\begin{equation}
	\frac{1}{\sqrt{-g}}\partial_\mu\left( {\sqrt{-g}}g^{{\mu}{\nu}}\partial_\nu{\Phi} \right)=0,
	\label{KG-wave-eq}
\end{equation}
where $g_{\mu\nu}$ is the metric and $g$ is its determinant. 

In a completely general spherically symmetric and static spacetime with a line element
\begin{equation}
	ds^2=-f(r)dt^2+g(r)^{-1}dr^2+r^2d\Omega^2,
	\label{spherical-le}
\end{equation} 
we apply the separation of variables
\begin{equation}
	\Phi(t,r, \theta, \phi) = Y_l(\theta, \phi)\Psi(t,r)/r,
	\label{seperate-variable}
\end{equation}
where $Y_l(\theta, \phi)$ are spherical harmonics with the multipole number $l=0,1,2,\dots$,
to obtain the QNM  wave equation
\beeq
\frac{\partial^2\Psi}{\partial t^2}+\left(-\frac{\partial^2}{\partial r_*^2}+V(r)\right)\Psi=0.
\label{WE-time}
\eneq  
In the above equation, $r_*$ is the tortoise coordinate linked to the radial coordinate according to
\beeq
dr_*=\frac{dr}{\sqrt{f(r)g(r)}},
\label{tortoise}
\eneq
and 
\beeq
V(r)=f(r) \frac{l(l+1)}{r^2}+ \frac{1}{2r} \frac{d}{dr} \left[ f(r)g(r) \right]
\label{eq-scalarV}
\eneq
is the Regge-Wheeler or QNM potential.  Since the fundamental QNM of geometric perturbations in a black hole spacetime has the multipole number $l=2$, in the rest of the paper, we will focus on scalar perturbations with $l=2$.

\section{Sagittarius A* Supermassive Black Hole}
\label{Sec:SgrBH}

As it is pointed out by \Mycite{BH-DMspike2}, a realistic model supported by the observational data for the Sgr A* supermassive black hole at the center of the Milky Way galaxy leads us to the following information.  The mass of this black hole  is $M_{\text{BH}} = 4.1 \times 10^6 M_\odot$.
For $\gamma_{\text{sp}} =9/4$ ($\gamma=0$), we have $R_{\text{sp}}\approx 0.91$ kpc and $\rho_{\text{sp}} \approx 1.39 \times 10^{-24}$ g cm$^{-3}$.  In terms of black hole parameters,  $R_{\text{sp}}\approx 2.32 \times 10^{9} r_{\text{BH}}$ and $\rho_{\text{sp}} \approx 1.26 \times 10^{-27} \rho_{\text{BH}}$, where 
\begin{equation}
	r_{\text{BH}} = \frac{2G M_\text{BH}}{c^2} \approx 2.95\left(\frac{M_{\text{BH}}}{M_\odot}  \right) \text{km}
	\label{}
\end{equation}
is the horizon radius and 
\begin{equation}
	\rho_{\text{BH}}=\frac{M_{\text{BH}}}{\frac{4}{3}\pi r_{\text{BH}}^3}
	\label{}
\end{equation}
is the mass density of the black hole.  
In the case of $\gamma_{\text{sp}} =7/3$ ($\gamma=1$), we have $R_{\text{sp}}\approx 0.235$ kpc and  
$\rho_{\text{sp}}\approx 8.00 \times 10^{-23}$ g cm$^{-3}$.  
In terms of black hole parameters, $R_{\text{sp}}\approx 6.00 \times 10^8 r_{\text{BH}}$ and $\rho_{\text{sp}} \approx 7.27 \times 10^{-26} \rho_{\text{BH}}$.  We also present this information in the table below where we include the values for $\alpha_\gamma$ and the total mass of the the spike, $M_{\text{total}}^{\text{sp}}$.

\vspace{0.5cm}
\begin{tabular}{cccccc}
	\multicolumn{6}{c}{Table I: DM Spike surrounding Sgr A* Supermassive Black Hole} \\  	
	\hline
	\vspace{-0.3cm}\\
	$\gamma_\text{sp}$ & $M_\text{BH}$ ($M_\odot$)& $\alpha_\gamma$ & $R_{\text{sp}}$ (kpc) &  $\rho_{\text{sp}}$ (g cm$^{-3}$) & $M_{\text{total}}^{\text{sp}}$ ($M_\odot$) \\ 
	\hline 
	$7/3$& $4.1 \times 10^{6}$  & $1.94$ & $0.235$ &  $8.00 \times 10^{-23}$ & $2.89\times 10^{8}$  \\ 
	$9/4$& $4.1 \times 10^{6}$ & $1.94$ & $0.910$ &  $1.39 \times 10^{-24}$ & $2.59\times 10^{8}$   \\ 
	\label{Table1}
\end{tabular} 
\normalsize

\Mycite{BH-DMspike2} also obtain upper bounds on $\rho_\text{sp}$ using the conditions that have to be satisfied everywhere outside the black hole horizon.  These conditions are

1. the metric determinant is always negative

2. $g_{\phi \phi}$ is always greater than zero, and

3.  $g_{rr}$ remains finite.
\newline
For $r_\text{b}=2 r_\text{BH}$, these upper bounds are calculated numerically in \Mycite{BH-DMspike2} for the two cases in Table I.  These bounds are:
\begin{equation}
	\gamma_{\text{sp}} =7/3,~~ R_{\text{sp}}= 0.235~ \mbox{kpc}: ~~\rho_{\text{sp}} < 2.37 \times 10^{-18} \mbox{g cm$^{-3}$}~.
	\label{eq-upperbound73}
\end{equation}
\begin{equation}
\gamma_{\text{sp}} =9/4,~~ R_{\text{sp}}= 0.91~\mbox{kpc}:~~ \rho_{\text{sp}} < 5.34 \times 10^{-19} \mbox{g cm$^{-3}$}~.~~
\label{eq-upperbound94}
\end{equation}

\begin{figure}[th!]
	\begin{center}
		\includegraphics[height=5.cm]{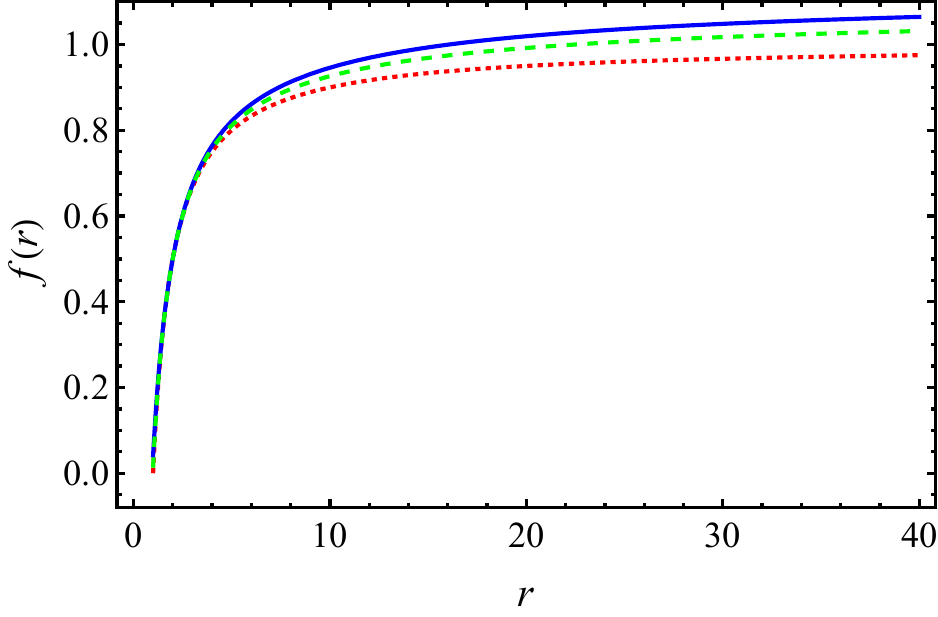}
		\includegraphics[height=5.cm]{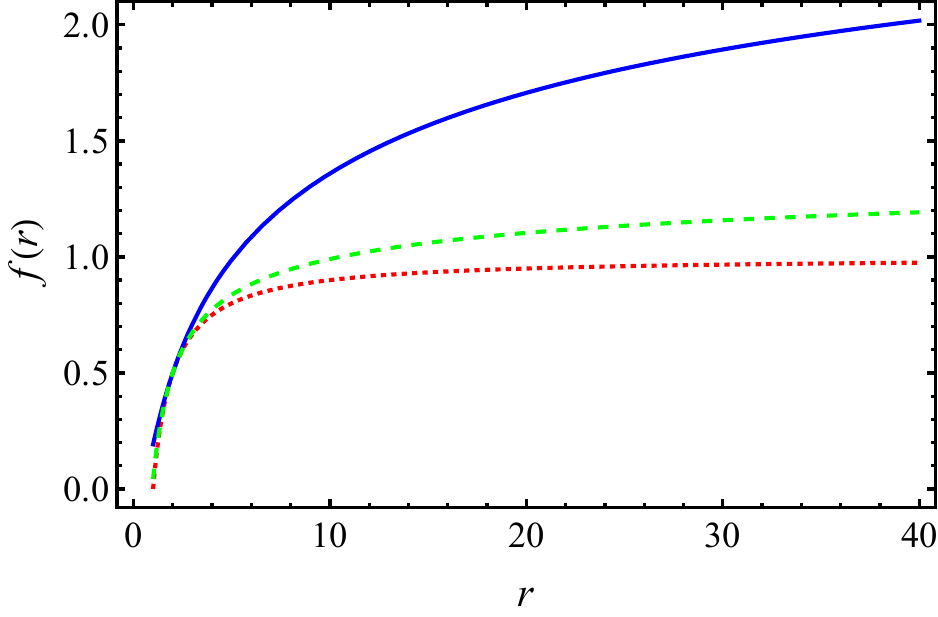}
	\end{center}
	\vspace{-0.7cm}
	\caption{\footnotesize Metric function $f$ versus radial coordinate (in terms of $r_\text{BH}$) for $\gamma_{\text{sp}} =7/3$, $r_\text{b}=2 r_\text{BH}$, and $R_\text{sp}=0.235$ kpc for the Sgr A* black hole surrounded by a DM spike. On the left,  $\rho_{\text{sp}}=6.7 \times 10^{-20}$ g cm$^{-3}$ ($\approx840$ times the expected value) and on the right,   $\rho_{\text{sp}}=4.7 \times 10^{-19}$ g cm$^{-3}$ ($\approx 6000$ times the expected value). In solid blue, we plot $f(r)$ given in Eq.\ (\ref{eq:f-7over3}).  In dashed green, we plot the function $f(r)$ given in Eq.\ (16) of \Mycite{BH-DMspike2}.  For comparison, we include the Schwarzschild metric function in dotted red.}
	\label{fig-metric-f}
\end{figure}

{To compare the metric function $f(r)$ obtained in this paper with the one provided by \Mycite{BH-DMspike2}, we plot both functions [Eq.\ (16) of \Mycite{BH-DMspike2} and our function provided in Eq.\ (\ref{eq:f-7over3})] in Figure \ref{fig-metric-f}, where all parameters are expressed in units of black hole parameters. The two functions differ significantly from each other for larger values of $\rho_\text{sp}$. This is presumably related to the fact that the authors in \Mycite{BH-DMspike2} derive an approximate metric function from Eq.\ (\ref{eq:alpha}), whereas ours is exact.}

\begin{figure}[th!]
	\begin{center}
		\includegraphics[height=5.3cm]{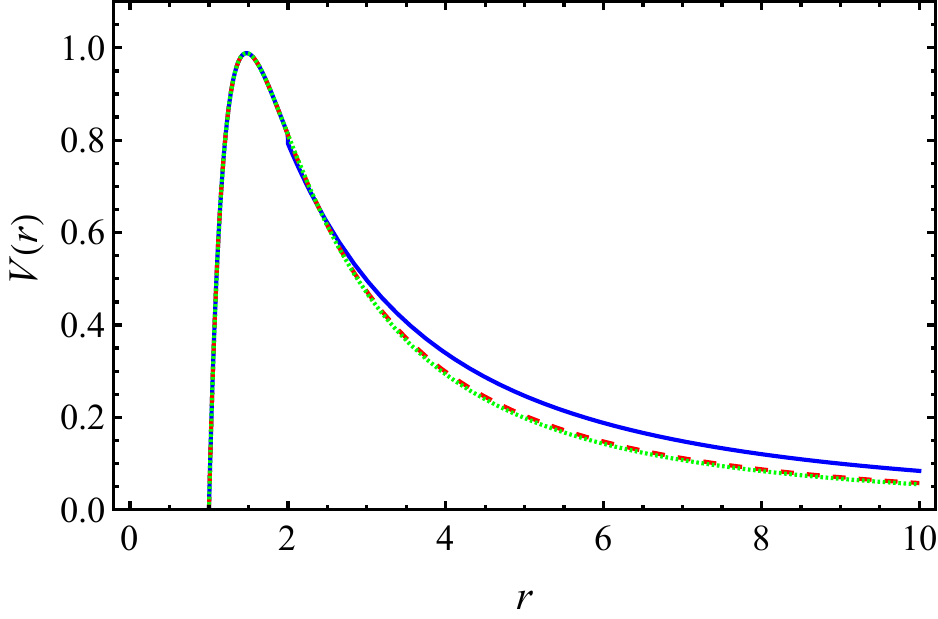}
	\end{center}
	\vspace{-0.7cm}
	\caption{\footnotesize Scalar QNM potential as a function of radial coordinate for $l=2$ and $\gamma_{\text{sp}} =7/3$ for the Sgr A* black hole surrounded by a DM spike. In dashed red, $\rho_{\text{sp}}=6.7 \times 10^{-20}$ g cm$^{-3}$ ($\approx 840$ times the expected value) and in solid blue, $\rho_{\text{sp}}=4.7\times 10^{-19}$ g cm$^{-3}$ ($\approx 6000$ times the expected value).  In both cases, $r_\text{b}=2 r_{\text{BH}}$ and $R_{\text{sp}}=0.235$ kpc.  For comparison, we include the Schwarzschild potential in dotted green. All our variables are expressed in terms of black hole parameters.}
	\label{fig-g1MW}
\end{figure}

To see how the DM spike influences the shape of the QNM potential given in Eq.\ (\ref{eq-scalarV}), we plot the potential for the case of $\gamma_\text{sp}=7/3$ in Figure \ref{fig-g1MW}.  A noticeable difference begins to appear when  $\rho_\text{sp}$ is roughly $840$ times bigger than the expected value presented in Table I. We also plot the the potential for $6000$ times bigger than the expected value of  $\rho_\text{sp}$.  All these density values are far less than the upper bound presented in Eq.\ (\ref{eq-upperbound73}).

\begin{figure}[th!]
	\begin{center}
		\includegraphics[height=5.3cm]{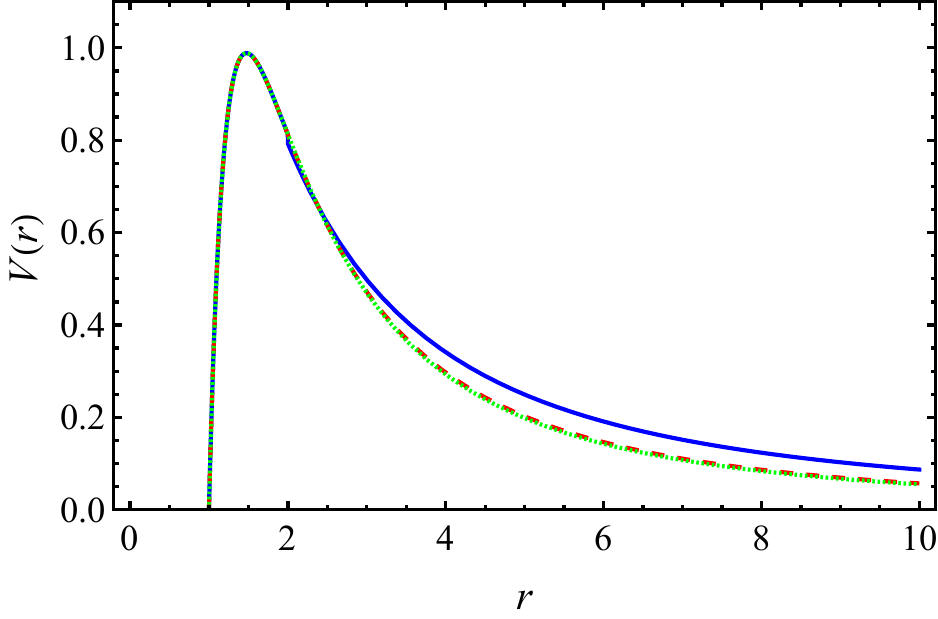}
	\end{center}
	\vspace{-0.7cm}
	\caption{\footnotesize Scalar QNM potential as a function of radial coordinate for $l=2$ and $\gamma_{\text{sp}} =9/4$ for the Sgr A* black hole surrounded by a DM spike. In dashed red, $\rho_{\text{sp}}=1.2 \times 10^{-20}$ g cm$^{-3}$ ($\approx 8400$ times the expected value) and in solid blue, $\rho_{\text{sp}}=1.2 \times 10^{-19}$ g cm$^{-3}$ ($\approx 84000$ times the expected value).  In both cases,  $r_\text{b}=2 r_{\text{BH}}$ and $R_{\text{sp}}=0.91$ kpc.  For comparison, we include the Schwarzschild potential in dotted green. All our variables are expressed in terms of black hole parameters.}
	\label{fig-g0MW}
\end{figure}

In Figure \ref{fig-g0MW}, we plot the potential (\ref{eq-scalarV})  for the case of $\gamma_\text{sp}=9/4$ .  A noticeable difference begins to appear when  $\rho_\text{sp}$ is roughly $8400$ times bigger than the expected value presented in Table I. We also plot the the potential for $84000$ times bigger than the expected value of  $\rho_\text{sp}$. As one can see, higher values of  $\rho_\text{sp}$ is required to observe noticeable change in the potential  for $\gamma_\text{sp}=9/4$ in comparison to the case of  $\gamma_\text{sp}=7/3$.  All these $\rho_\text{sp}$ values are still less than the upper bound presented in Eq.\ (\ref{eq-upperbound94}).

To generate the ringdown waveform, we numerically solve the time-dependent wave equation (\ref{WE-time}) using the initial data
\beeq
\Psi(r_*,0)={\cal A} \exp \left(- \frac{(r_*-\bar{r}_{*})^2}{2\sigma^2} \right),~  \partial_t \Psi|_{t=0}=-\partial_{r_*} \Psi(r_*, 0)~,
\label{GaussianWave}
\eneq  
where we use $\sigma=1 ~r_\text{BH}$, $\bar{r}_*=-40 ~r_\text{BH}$, and ${\cal A}=10 ~r_\text{BH}^{-2}$.  We choose the observer to be located at $r_*=90 ~r_\text{BH}$. In all the cases studied here,  the height of the QNM potential at $r_*=90 ~r_\text{BH}$, which is inside the spike region, is small ($\lessapprox 10^{-3} r_\text{BH}^{-2}$) compared to the peak. Therefore, we do not expect a significant difference in the results if the observer is further away.  

\begin{figure}[th!]
	\begin{center}
		\includegraphics[height=5.cm]{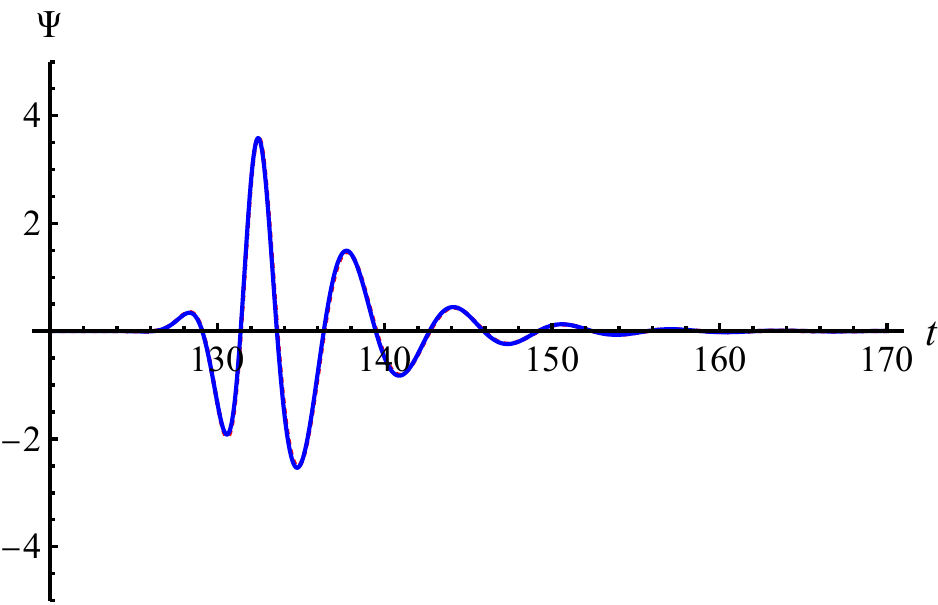}
		\includegraphics[height=5.cm]{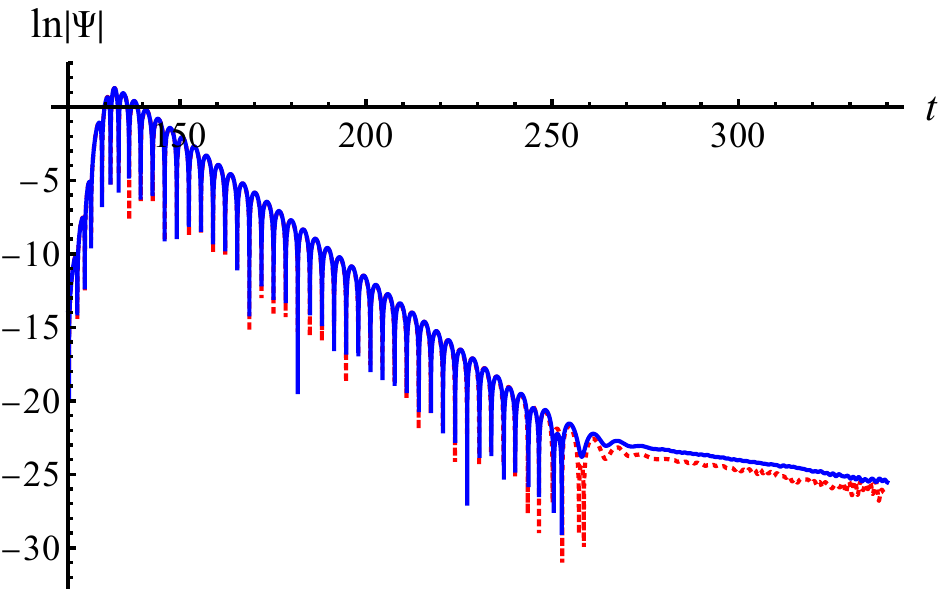}
	\end{center}
	\vspace{-0.7cm}
	\caption{\footnotesize In solid blue, ringdown waveform $\Psi$ (left) and $\ln |\Psi|$ (right) as a function of time for $l=2$, $\gamma_{\text{sp}}=7/3$, $R_{\text{sp}}=0.235$ kpc, and $\rho_{\text{sp}}=6.7 \times 10^{-20}$ g cm$^{-3}$ ($\approx 840$ times the expected value) for the Sgr A* black hole surrounded by a DM spike.  For comparison, we include the Schwarzschild ringdown waveform  in dotted red. All our variables are expressed in terms of black hole parameters.}
	\label{fig-SgrGraph100}
\end{figure}

\begin{figure}[th!]
	\begin{center}
		\includegraphics[height=5.cm]{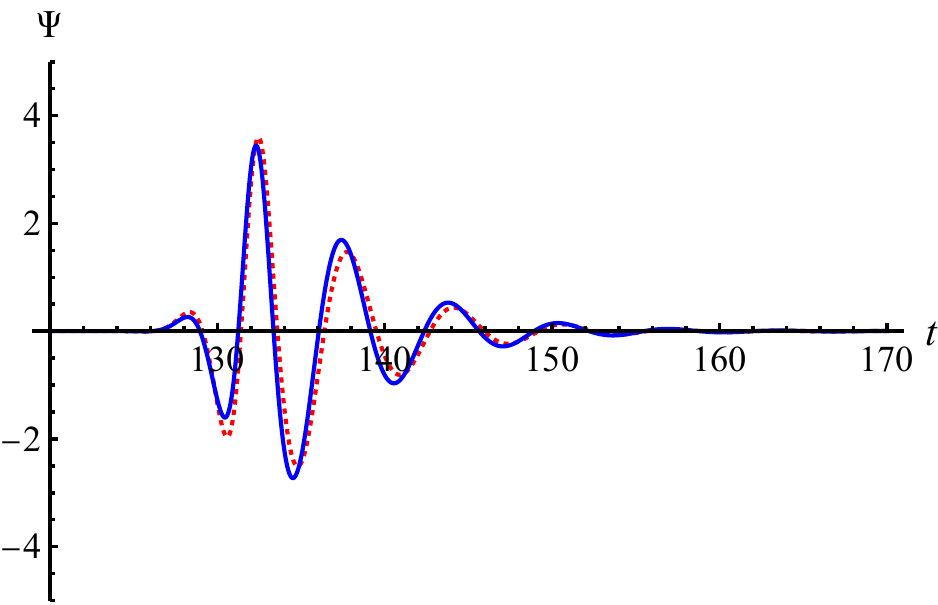}
		\includegraphics[height=5.cm]{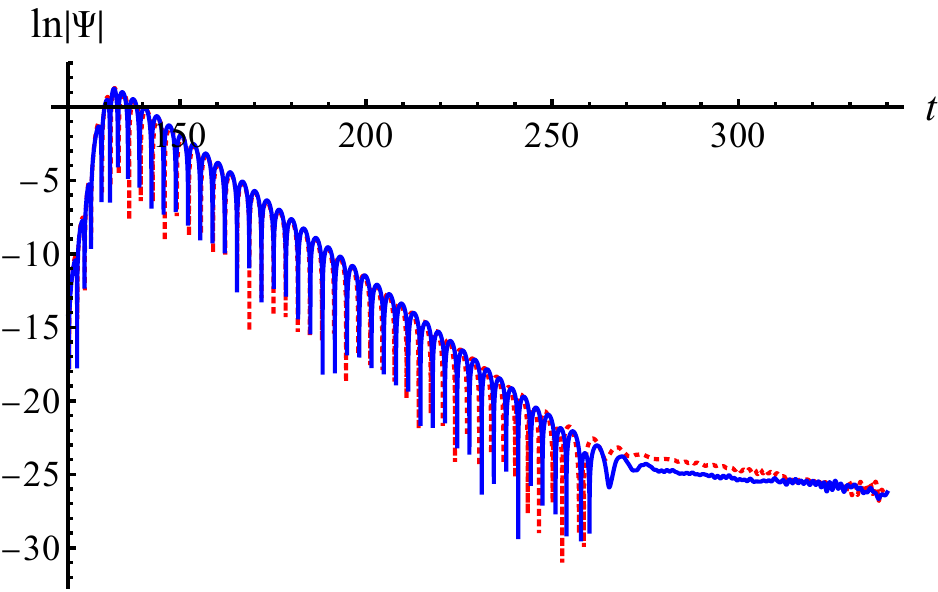}
	\end{center}
	\vspace{-0.7cm}
    \caption{\footnotesize In solid blue, ringdown waveform $\Psi$ (left) and $\ln |\Psi|$ (right) as a function of time for $l=2$, $\gamma_{\text{sp}}=7/3$, $R_{\text{sp}}=0.235$ kpc, and $\rho_{\text{sp}}=4.7 \times 10^{-19}$ g cm$^{-3}$ ($\approx 6000$ times the expected value) for the Sgr A* black hole surrounded by a DM spike.  For comparison, we include the Schwarzschild ringdown waveform  in dotted red. All our variables are expressed in terms of black hole parameters.}
	\label{fig-SgrGraph700}
\end{figure}

To carry out the calculations, we use the built-in {\em Mathematica} commands for solving partial differential equations.   We check the accuracy of our results by computing the waveforms for Schwarzschild and comparing them to known results.
The resulting ringdown waveforms for the potentials shown in Figure \ref{fig-g1MW} are plotted in Figures \ref{fig-SgrGraph100} and \ref{fig-SgrGraph700}.

We also extract the first ($n=0$) QNM frequency from these waveforms using the Prony method (\Mycitet{Prony}), a numerical procedure that fits N data points by as many purely damped exponentials as necessary. 
To test the Prony method, we first calculate the QNM of scalar perturbations for the Schwarzschild case for $l=2$.  The result is $0.967442-0.193137i$, which is in good agreement with the value $0.967284 -0.193532 i$ found using the continued fraction method (\Mycitet{DGMK}).  We find $0.966059-0.193151i$ for the case presented in Figure \ref{fig-SgrGraph100} and $0.956060-0.194191i$ for the case in Figure \ref{fig-SgrGraph700}.  It is clear that as we increase $\rho_{\text{sp}}$ (and consequently $M_\text{total}^\text{sp})$, the real part (oscillation frequency) of the QNM decreases and the imaginary part (damping)  increases.

\newpage

\section{M87 Supermassive Black Hole}
\label{Sec:M87}

In this section, we use the data provided by \Mycite{M87data}.  The authors
use $M_{\text{BH}}=6.4 \times 10^9 M_\odot$ ($r_{\text{BH}}=6 \times 10^{-4}$ pc) and   
fix the initial halo power law parameter $r_0$ to be $20$ kpc (as for the Milky Way). They assume $\alpha_\gamma=0.1$.  The authors then determine $\rho_0\approx 2.5$ GeV cm$^{-3}$ for $\gamma=1$ ($\gamma_\text{sp}=7/4$) based on the observational data provided in \Mycite{M87-observation}.   With these values, we can use Eq.\ (\ref{eq-spike-parameters}) to  evaluate $R_{\text{sp}}\approx 0.219$ kpc and   $\rho_{\text{sp}}\approx 4.10 \times 10^{-22}$ g cm$^{-3}$. 
In terms of black hole parameters, we have $R_{\text{sp}}\approx  3.59\times 10^5 ~r_{\text{BH}}$ and $\rho_{\text{sp}}=9.1 \times 10^{-19} \rho_{\text{BH}}$.  
The DM density profile in different regions around the M87 black hole is summarized in \Mycite{M87data} as\footnote{Since we calculate the waveforms at radii far from the horizon, but still inside the spike, the form of the density outside the spike is irrelevant.}
\begin{equation}
	\rho(r)=
	\left\{ {\begin{array}{cc}
			0, & r \le r_\text{BH} \\
			\rho_\text{DM}^\text{sp}(r), &  r_\text{BH} \le r < R_{\text{sp}} \\
			\rho_0  \left( \frac{r}{r_0} \right)^{-\gamma}  \left( 1+\frac{r_0}{r} \right)^{-2}  & r \ge R_{\text{sp}} \\
	\end{array} } \right. 
	\label{eq-LacroixDensity}
\end{equation}

It is important to note that in \Mycite{BH-DMspike2}  $\alpha_\gamma \approx 1.94$. If we choose this value for $\alpha_\gamma$, together  with the $r_0$ and $\rho_0$ values mentioned above, we can use Eqs.\ (\ref{eq-spike-parameters}) and (\ref{eq-SpikeMass}) to calculate the total mass of the DM spike to be  $M_{\text{total}}^{\text{sp}}=4.54\times 10^{11}M_\odot$.  We can also use Eq.\ (\ref{eq-LacroixDensity}) to calculate the mass of the DM halo outside the spike region ($r \ge R_{\text{sp}}$) to obtain
\begin{equation}
	M_\text{total}^\text{halo}=\int_{R_{\text{sp}}}^{50 \text{kpc}} 4 \pi r^2 \rho_0  \left( \frac{r}{r_0} \right)^{-\gamma}  \left( 1+\frac{r_0}{r} \right)^{-2} dr = 3.48 \times 10^{12}M_\odot .
	\label{}
\end{equation}
Adding the masses of the DM spike and black hole to the mass of the halo, we find the total mass of $3.94 \times 10^{12}M_\odot$.
This mass is within an acceptable range based on the observational data that estimate the total mass of M87 within $50$ kpc radius to be $6\times 10^{12}M_\odot$ (\Mycitet{M87-observation}).  
Also note that the spike mass, for the $\alpha_\gamma \approx 1.94$ case, is $100$ times bigger than the mass of the M87 black hole.  A similar DM spike to black hole mass ratio holds for the values presented in \Mycite{BH-DMspike2} for the Sgr A* case.  Therefore, to have a sensible comparison between the Sgr A* and M87 cases, one should use the same  $\alpha_\gamma$, and consequently the same spike to black hole mass ratio, for both. 
We summarize the parameters for M87 in Table II, where we also include the parameters for $\alpha_\gamma = 1.94$.
{To the best of our knowledge there is no observational bound on $\alpha_\gamma$. Both values that we use in Table II are consistent with the data provided in \Mycite{M87-observation}.}

\vspace{0.5cm}
\begin{tabular}{cccccc}
	\multicolumn{6}{c}{Table II: DM Spike surrounding {M87}  Supermassive Black Hole} \\  	
	\hline
	\vspace{-0.3cm}\\
	$\gamma_\text{sp}$ & $M_\text{BH}$ ($M_\odot$)& $\alpha_\gamma$ & $R_{\text{sp}}$ (kpc) &  $\rho_{\text{sp}}$ (g cm$^{-3}$) & $M_{\text{total}}^{\text{sp}}$ ($M_\odot$) \\ 
	\hline 
	$7/3$& $6.4 \times 10^{9}$  & $0.1$ & $0.219$ &  $4.10 \times 10^{-22}$ & $1.21\times 10^{9}$  \\ 
	$7/3$& $6.4 \times 10^{9}$ & $1.94$ & $4.26$ &  $2.12 \times 10^{-23}$ & $4.54\times 10^{11}$   \\ 
	\label{Table2}
\end{tabular} 
\normalsize

\begin{figure}[th!]
	\begin{center}
		\includegraphics[height=5.3cm]{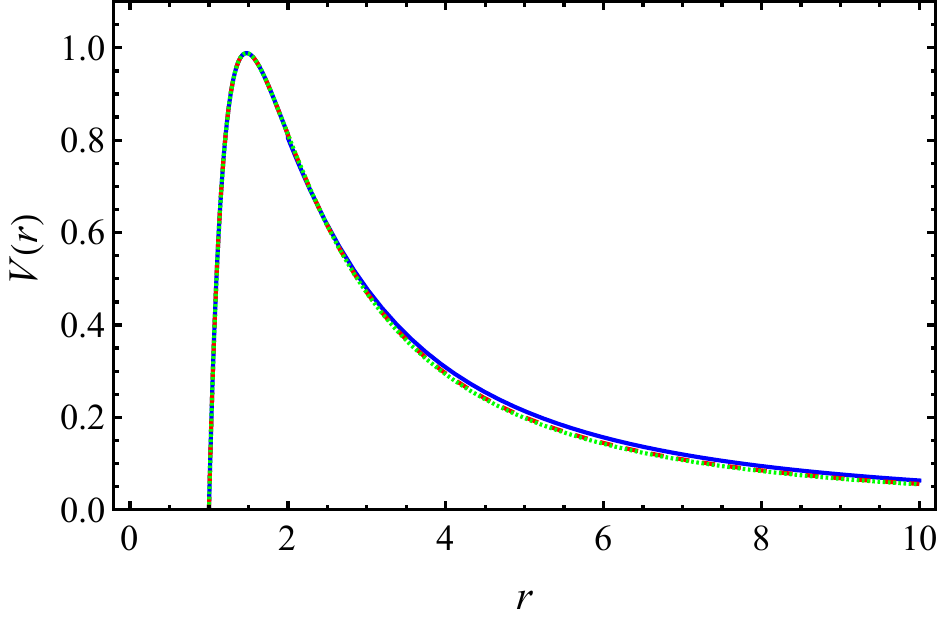}
	\end{center}
	\vspace{-0.7cm}
	\caption{\footnotesize Scalar QNM potential as a function of radial coordinate for $l=2$ and  $\gamma_{\text{sp}} =7/3$ for the M87 black hole surrounded by a DM spike. In dashed red,  $\rho_{\text{sp}}=3.4 \times 10^{-19}$ g cm$^{-3}$ ($\approx 840$ times the expected value) and in solid blue, $\rho_{\text{sp}}=2.4 \times 10^{-18}$ g cm$^{-3}$ ($\approx 6000$ times the expected value).  In both cases, $r_\text{b}=2 r_{\text{BH}}$ and $R_{\text{sp}}=0.219$ kpc.  For comparison, we include the Schwarzschild potential in dotted green.  
	All our variables are expressed in terms of black hole parameters.}
	\label{fig-potentialM87}
\end{figure}

To compare the shape of the black hole QNM potential in the presence of the DM spike with the Schwarzschild case, in Figure \ref{fig-potentialM87}, we plot the potential for the case of $\alpha_\gamma=0.1$.  A noticeable difference begins to appear when  $\rho_\text{sp}$ is roughly $840$ times bigger than the expected value presented in Table II. We also plot the potential for $6000$ times bigger density than the expected value. 

\begin{figure}[th!]
	\begin{center}
		\includegraphics[height=5.3cm]{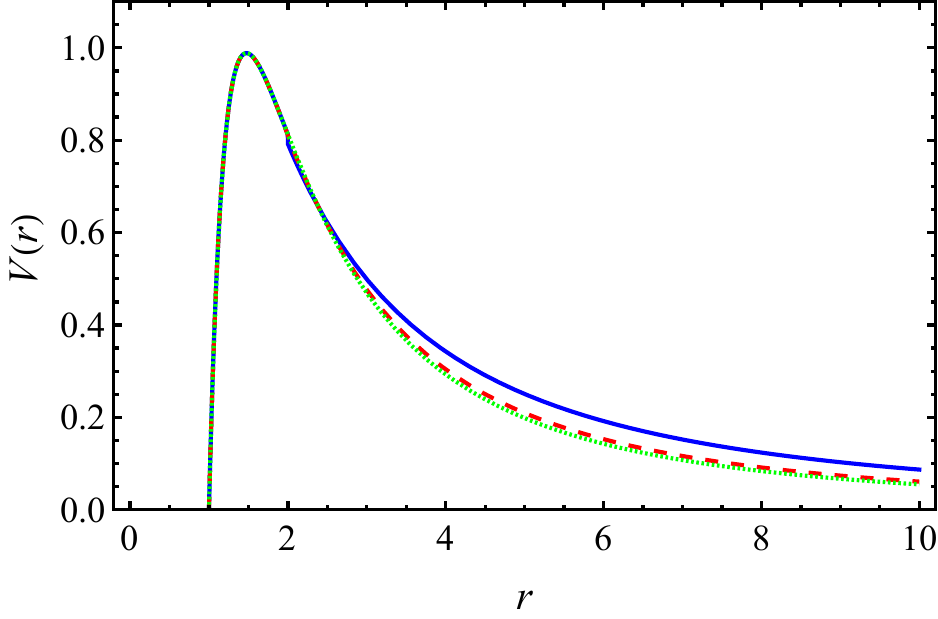}
	\end{center}
	\vspace{-0.7cm}
	\caption{\footnotesize Scalar QNM potential as a function of radial coordinate for $l=2$ and  $\gamma_{\text{sp}}=7/3$ for the M87 black hole surrounded by a DM spike. In dashed red, $\rho_{\text{sp}}=1.8 \times 10^{-21}$ g cm$^{-3}$ ($\approx 84$ times the expected value) and in solid blue, $\rho_{\text{sp}}=6.8 \times 10^{-21}$ g cm$^{-3}$ ($\approx 320$ times the expected value).  In both cases, $R_{\text{sp}}=4.26$ kpc.  
		For comparison, we include the Schwarzschild potential in dotted green.
	All our variables are expressed in terms of black hole parameters.}
	\label{fig-potentialM87-194}
\end{figure}

As it was discussed, for a sensible comparison between the Sgr A* and M87 cases, one should use the same value for $\alpha_\gamma$. Therefore, in  Figure \ref{fig-potentialM87-194}, we plot a similar graph to Figure \ref{fig-g1MW} for the M87 case when $\alpha_\gamma=1.94$.   A noticeable difference in the potential begins to appear when  $\rho_\text{sp}$ is roughly $84$ times bigger than the expected value presented in Table II. Note that in Figure \ref{fig-g1MW}, a noticeable difference in the potential appears only when $\rho_\text{sp}$ is roughly $840$ times bigger.   In  Figure \ref{fig-potentialM87-194}, we also plot the potential for $\rho_\text{sp}$ with a value of  $320$ times bigger than the expected value, which is more or less has the same impact on the potential as the $6000$ times bigger  $\rho_\text{sp}$ in the Sgr A* case. This shows it is easier to detect the DM spike in M87 using the ringdown waveform in comparison to the Sgr A* black hole assuming the spike to black hole mass ratio is roughly constant for both galaxies.

\begin{figure}[th!]
	\begin{center}
		\includegraphics[height=5.cm]{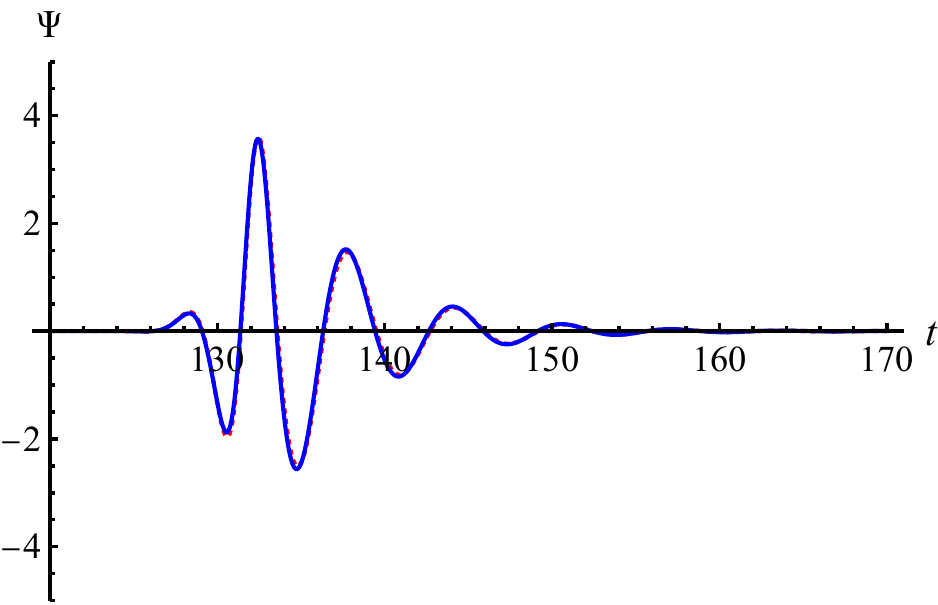}
		\includegraphics[height=5.cm]{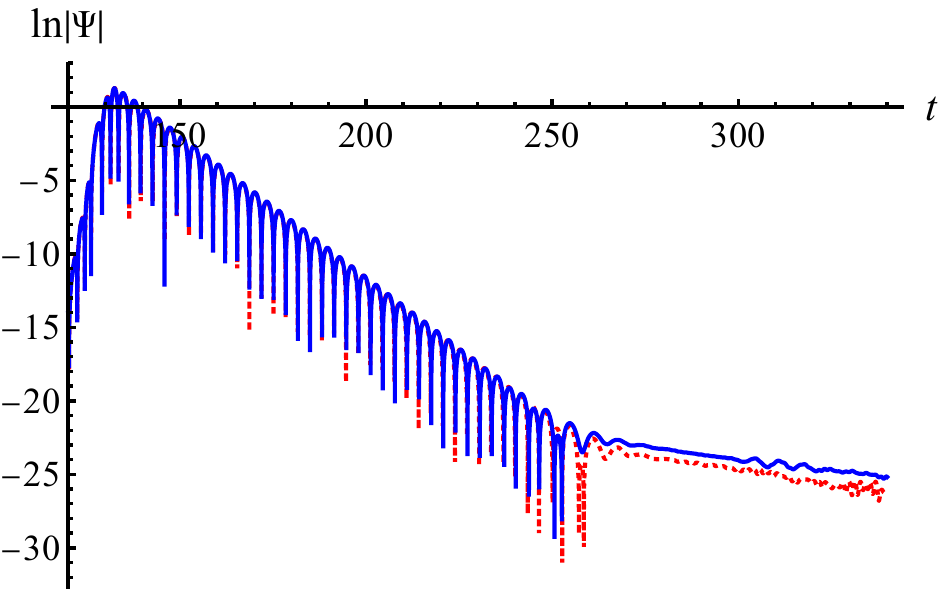}
	\end{center}
	\vspace{-0.7cm}
	\caption{\footnotesize In solid blue, ringdown waveform $\Psi$ (left) and $\ln |\Psi|$ (right) as a function of time for $l=2$, $\gamma_{\text{sp}}=7/3$, $R_{\text{sp}}=4.26$ kpc, and $\rho_{\text{sp}}=1.8\times 10^{-21}$ g cm$^{-3}$ ($\approx 84$ times the expected value) for the M87 black hole surrounded by a DM spike.  For comparison, we include the Schwarzschild ringdown waveform in dotted red. All our variables are expressed in terms of black hole parameters.}
	\label{fig-M87-ringdown-10}
\end{figure}
\begin{figure}[th!]
 	\begin{center}
 		\includegraphics[height=5.cm]{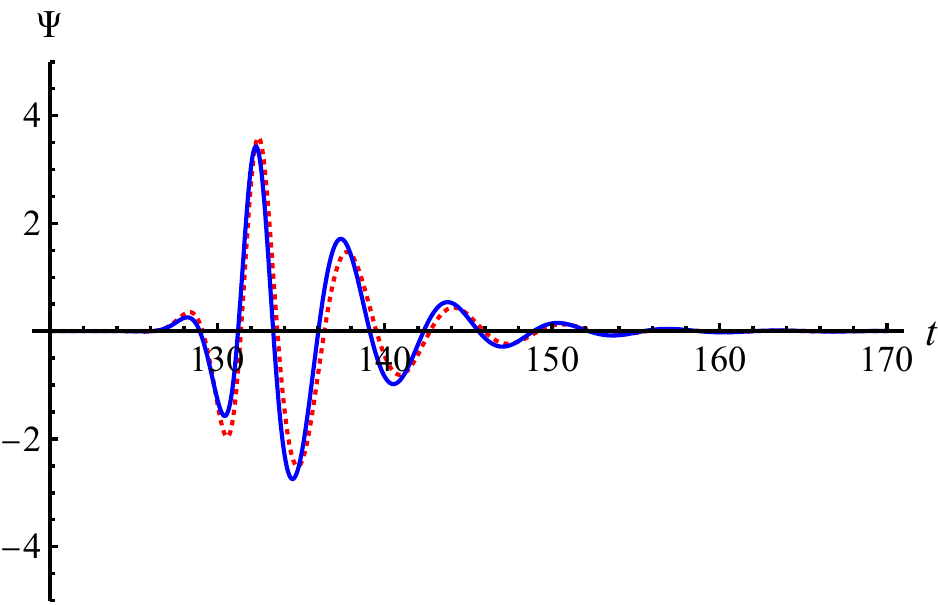}
 		\includegraphics[height=5.cm]{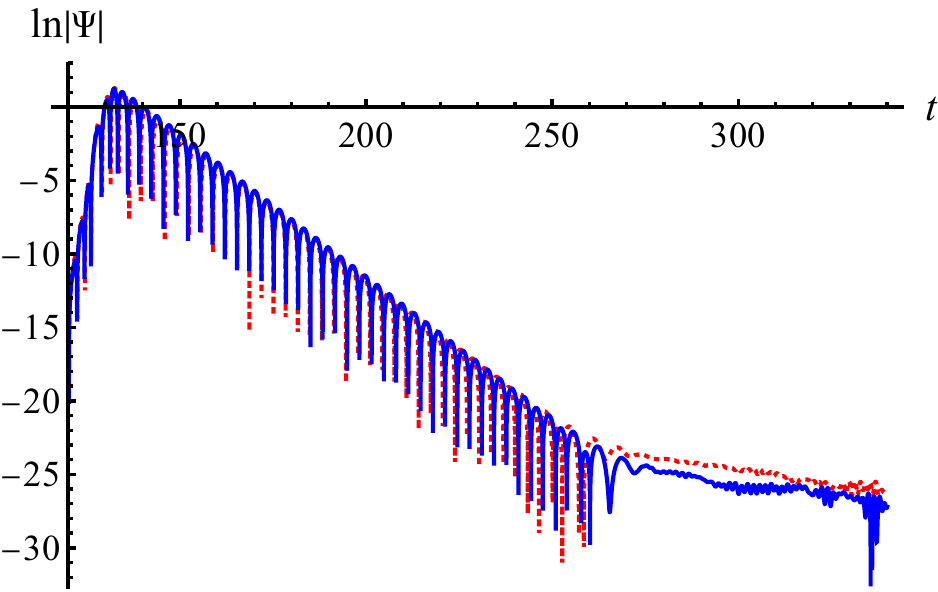}
 	\end{center}
 	\vspace{-0.7cm}
 	\caption{\footnotesize In solid blue, ringdown waveform $\Psi$ (left) and $\ln |\Psi|$ (right) as a function of time for $l=2$, $\gamma_{\text{sp}}=7/3$, $R_{\text{sp}}=4.26$ kpc, and $\rho_{\text{sp}}=6.8 \times 10^{-21}$ g cm$^{-3}$ ($\approx 320$ times the expected value) for the M87 black hole surrounded by a DM spike.  For comparison, we include the Schwarzschild ringdown waveform in dotted red. All our variables are expressed in terms of black hole parameters.}
 	\label{fig-M87-ringdown-38}
 \end{figure}
We generate the ringdown waveforms for the potentials shown in Figure \ref{fig-potentialM87-194}.  These waveforms are plotted in Figures \ref{fig-M87-ringdown-10} and \ref{fig-M87-ringdown-38}.  
We use the Prony method (\Mycitet{Prony}) to extract the first ($n=0$) QNM frequency from the waveforms shown in Figures \ref{fig-M87-ringdown-10} and \ref{fig-M87-ringdown-38}.  
We find $0.964669-0.193195i$ for the case presented in Figure \ref{fig-M87-ringdown-10} and $0.955183-0.194375i$ for the case in Figure \ref{fig-M87-ringdown-38}.  It is clear that as we increase $\rho_{\text{sp}}$, the real part (oscillation frequency) of the QNM decreases and the imaginary part (damping)  increases.


\section{Summary and Conclusion}
\label{Sec:conclusions}

We have used the TOV equations to construct the spacetime metric representing a black hole surrounded by a perfect fluid DM spike.  Following previous work, we assumed a power law density for the DM spike, which was therefore completely specified by three independent parameters: the power law exponent $\gamma_\text{sp}$, the radius $R_\text{sp}$ and spike density $\rho_\text{DM}^\text{sp}$, the latter two chosen to lie at the outer edge of the spike. These in turn determine the total mass of the spike. Given the black hole mass, the TOV equations then determined uniquely the metric of the spacetime containing the spike. With this metric, we were able to calculate the ringdown waveform of the gravitational waves associated with black hole perturbations, as well as the real and imaginary parts of the lowest damping QNM.

The main features that emerge from our analysis were:
\begin{itemize}
	\item The pressure inside the spike is negligible in all the cases we studied.
	\item The presence of the DM spike modifies the ringdown waveform.  More specifically, it decreases the real part (oscillation frequency) of the least damped QNM and increases its imaginary part (damping).
	\item For Sgr A*  the parameters have to be pushed well beyond the accepted ranges in order to produce significant differences from Schwarzschild ringdown waveform. The prospects of detection are therefore remote.
	\item For M87, the parameters are less known, but there is an observational  bound on the total mass within $50$ kpc of the center, which in turn provides an upper bound on the spike mass.  We find that while the departures from Schwarzschild for the ringdown waveforms are significantly greater for  M87 than for Sgr A*, the spike mass needs to be an order of magnitude or two above the proposed upper bound in order to have hopes of detecting it with current gravitational wave technology.
	\item Our results also suggest that if the ratio of DM spike mass to black hole mass is roughly constant for galactic black holes, greater mass black holes require smaller spike densities in order to yield potentially observable signals.
\end{itemize}

We conclude  that a significant gravitational wave detection associated with perturbations of a supermassive black hole more massive than the M87 black hole might provide the means to detect the presence of a DM spike or at least put a model dependent bound on its parameters. This suggests that the effects of DM spikes on the ringdown waveforms of supermassive black holes are worthy of further study.

{In this paper, we focused on static spherically symmetric black holes.  It would be interesting to see what happens if spin is introduced.  \Mycite{Ferrer} argued that the spike will be enhanced by the presence of spin.  We therefore expect that the inclusion of the black hole spin will improve our results in terms of observational viability. This is currently under investigation.}

\hspace{5cm}

\leftline{\bf Acknowledgments}
We thank Michael Green for sharing with us the code to create ringdown waveforms and helping us with the Prony method. 
G.K.\ gratefully acknowledges that this research was supported in
part by a Discovery Grant from the Natural Sciences and Engineering Research Council of
Canada.



\def\jnl#1#2#3#4{{#1}{\bf #2} #3 (#4)}

\def\Zphys{{Z.\ Phys.} }
\def\jssc{{J.\ Solid State Chem.\ }}
\def\jpsJ{{J.\ Phys.\ Soc.\ Japan }}
\def\ptps{{Prog.\ Theoret.\ Phys.\ Suppl.\ }}
\def\PTP{{Prog.\ Theoret.\ Phys.\  }}
\def\LNC{{Lett.\ Nuovo.\ Cim.\  }}

\def\JMP{{J. Math.\ Phys.} }
\def\NPB{{Nucl.\ Phys.} B}
\def\NP{{Nucl.\ Phys.} }
\def\PLB{{Phys.\ Lett.} B}
\def\PL{{Phys.\ Lett.} }
\def\PRL{Phys.\ Rev.\ Lett.\ }
\def\PRA{{Phys.\ Rev.} A}
\def\PRB{{Phys.\ Rev.} B}
\def\PRD{{Phys.\ Rev.} D~}
\def\PR{{Phys.\ Rev.} }
\def\PRe{{Phys.\ Rep.} }
\def\AP{{Ann.\ Phys.\ (N.Y.)} }
\def\RMP{{Rev.\ Mod.\ Phys.} }
\def\ZPC{{Z.\ Phys.} C}
\def\SCI{Science}
\def\CMP{Comm.\ Math.\ Phys. }
\def\MPLA{{Mod.\ Phys.\ Lett.} A}
\def\IJMPA{{Int.\ J.\ Mod.\ Phys.} A}
\def\IJMPB{{Int.\ J.\ Mod.\ Phys.} B}
\def\cmp{{Com.\ Math.\ Phys.}}
\def\JPA{{J.\  Phys.} A}
\def\CQG{Class.\ Quant.\ Grav.~}
\def\ATMP{Adv.\ Theoret.\ Math.\ Phys.~}
\def\AJP{Am.\ J.\ Phys.~}
\def\PRSA{{Proc.\ Roy.\ Soc.\ Lond.} A }
\def\ibid{{ibid.} }
\vskip 1cm


\makeatletter
\renewcommand\@biblabel[1]{}
\makeatother


\end{document}